%% file: ms.tex
\documentclass[]{aa}
\usepackage{graphicx,natbib}
\usepackage{lipsum}
\usepackage{latexsym,amsmath,tabularx,amssymb}
\usepackage{mathrsfs}
\usepackage[latin1]{inputenc}
\usepackage[T1]{fontenc}
\bibpunct{(}{)}{;}{a}{}{,}
\usepackage[varg]{txfonts}

\include{def}
\usepackage{bm}
\usepackage{units}
\usepackage{color}
\usepackage{url}

\begin{document}

\title{Using Deep Neural Networks to compute the mass of forming planets}

\author{Y. Alibert \inst{1}, J. Venturini \inst{2}}
\offprints{Y. Alibert}
\institute{Physikalisches Institut  \& NCCR PlanetS, Universitaet Bern, CH-3012 Bern, Switzerland, 
\and
{International Space Science Institute, CH-3012 , Bern, Switzerland} \\
        \email{yann.alibert@space.unibe.ch,venturini@issibern.ch}}

\abstract
{Computing the mass of planetary envelopes and the critical mass beyond which planets accrete gas in a runaway fashion is important when studying planet formation, in particular for planets up to the Neptune mass range. This computation requires in principle solving a set of differential equations, the internal structure equations, for some boundary conditions (pressure, temperature in the protoplanetary disk where a planet forms, core mass and accretion rate of solids by the planet). Solving these equations in turn proves being time consuming and sometimes numerically unstable.}
{The aim is to provide a way to approximate the result of integrating the internal structure equations for a variety of boundary conditions.}
{We compute a set of planet internal structures for a very large number (millions) of boundary conditions, considering two opacities, namely the ISM one and a reduced one. This database is then used to train
Deep Neural Networks in order to predict the critical core mass as well as the mass of planetary envelopes as a function of the boundary conditions.}
{We show that our neural networks provide a very good approximation (at the level of percents) of  the result obtained by solving interior structure equations,  but with a much smaller required computer time. 
The difference with the real solution is much smaller than the one obtained using some analytical formulas available in the literature which at best only provide the correct order of magnitude. {We compare the results of the DNN with other popular machine learning methods (Random Forest, Gradient Boost, Support Vector Regression) and show that the DNN outperforms these methods by a factor of at least two.}}
{We show that some analytical formulas that can be found in various papers can severely overestimate the mass of planets, therefore predicting the formation of planets in the Jupiter-mass regime instead of the Neptune-mass regime. The python tools that we provide allow to compute the critical mass and the mass of planetary envelopes in a variety of cases, 
without having to solve the internal structure equations. These tools can easily replace the aforementioned analytical formulas, while providing much more accurate results.} 


\authorrunning{Y. Alibert  and J. Venturini}
\titlerunning{DNNs to compute the mass of forming planets}

\maketitle

\section{Introduction}
\label{sec:introduction}

Understanding the formation of planets (from terrestrial to gas giants) requires the development of theoretical models whose outcome can be compared with observations {\citep[e.g,][]{Mord09,Alibert19}}. 
One primary outcome of planet formation models is the mass and internal structure (in particular mean density) of planets, as these two quantities can be obtained from radial velocity and transit observations
of extrasolar planets. 

In the core accretion model {\citep[e.g.,][]{BP86, P96,Hubi05}}, the mass growth of planets results from two phenomena. The first is the accretion of solids \citep[{planetesimals} or pebbles, e.g][]{Alibert13,Ormel17},
which depends on the properties of the accreted solids, the planetary mass, and the disc thermodynamical properties \citep{Fortier13, OrmelK10}. The second is the accretion of gas which
occurs in two regimes \citep[see][for a review]{HelledPPVI}. 
In the first regime, when the planet is small, the planetary gas envelope extends until the protoplanetary disc, at a radius that is generally between the Bondi and Hill radius \citep{Lissauer09}.
The planet is said to be attached to the disc, and the accretion rate of gas depends on the cooling efficiency of the planet. Indeed, planets must cool in order to accrete gas  \citep{Lee15}, 
and the mass of the gas envelope in this regime must be computed by solving the equations for planetary interior structure \citep{Ikoma00,Venturini15}. 
When the mass of the planetary core is larger than the so-called critical mass, the radiative loss at the surface of the planet cannot be compensated for by the accretion energy of solids {\citep{mizuno80}}, 
and the planet starts to contract, in a runaway fashion\footnote{This does not imply that the accretion rate of gas is large, but that it is \textit{accelerating}}. {\citep{BP86, Ikoma00}}

For sub-critical planets (when the core mass is smaller than the critical mass), the computation of the planetary envelope requires solving a set of  differential equations (see Sect. \ref{internal_structure}).
Although standard procedure, solving these internal structure equations can require a non-negligible amount of computer time, 
and lead in some cases (e.g. close to the critical core) to some numerical instability. For this reason, some authors have developed fitting formula which allows estimating 
the envelope mass as a function of different parameters (e.g. opacity, core mass, etc.), as well as some analytical approximation of the critical core mass.
For example, \citet[][hereafter I00]{Ikoma00}, showed that the gas accretion rate is given by:
\begin{equation}
t_{\rm gas} = 10^b \left( {M_{\rm planet} \over \mearth } \right)^{-c} \left( { \kappa \over 1 \,  {\rm cm^{2}/g } }  \right) ,
\label{TKH}
\end{equation}
where $b=8$ and $c=2.5$. $M_{\rm planet}$  is the planetary mass and $\kappa$ the envelope opacity. This formula was derived in the absence of solid accretion, but the authors showed that the accretion timescale in the presence of a constant solid accretion follows a similar
law, the pre-factor being multiplied by six. Certainly, the accretion energy of solids provides some thermal support that slows down the accretion of gas. 

Using the same calculations, \citet[][hereafter IL04]{IdaLin04} showed that  the critical core mass can be approximated as:
\begin{equation}
M_{\rm crit} = 10 \left( { \dot{M}_{\rm core} \over 10^{-6} \mearth {\rm yr}^{-1} } \right)^{0.2-0.3} \left( { \kappa \over 1 \, {\rm cm^2/g} } \right)^{0.2-0.3} \mearth,
\label{Mcrit_IL04}
\end{equation}
where $\dot{M}_{\rm core}$ is the accretion rate of solids.

More recently, based on the results from \citet{PY14}, \citet[][hereafter B15]{Bitsch15b} used an accretion rate of gas given by:
\begin{eqnarray}
\dot{M}_{\rm gas} & = & 0.00175 f^{-2} \left( {\kappa \over 1 {\rm cm^2/g}}\right)^{-1} \left( { \rho _c \over {\rm 5.5 g/cm^3} } \right)^{-1/6} \left( { M_{\rm core} \over \mearth } \right)^{11/3}  \nonumber \\
&   &  \left( {M_{\rm env} \over 0.1 \mearth } \right)^{-1} \left( {T \over 81 K} \right)^{-0.5} {\mearth \over {\rm Myr}},
\label{B15}
\end{eqnarray}
where $\kappa$ is the opacity in the envelope, $M_{\rm core}$ and $M_{\rm env}$ are the core and envelope mass respectively, $ \rho _c$ is the core density, 
$T$ the temperature at the planet's location in the disk, and $f$ is a numerical factor of the order of 0.2.  
This formula is in principle valid in the absence of any solid accretion, but following the argument of I00, this accretion rate can be used in the case of a constant sold accretion
rate, by reducing factor $f$ from $0.2$ down to $\sim 0.033$. 

For super-critical planets, the accretion rate of gas can be approximated using the Kelvin-Helmoltz  timescale  $t_{\rm KH} \sim t_{\rm gas}$ as:
\begin{equation}
\dot{M}_{\rm gas} ={ M_{\rm planet} \over t_{\rm KH}} , 
\end{equation}
 $t_{\rm KH}$ itself depending on the planetary mass  (see Eq. \ref{TKH}, I00, IL04).

The use of these simplified formulas, although convenient from the numerical point of view, is  questionable when computing 
the formation of low mass planets, in the Super-Earth up to the Neptune mass range. Indeed, as the envelope mass of these planets is small (smaller than the core mass), an error of a couple of Earth masses can notably change their radius and other properties like their habitability \citep{Alibert14}.  This was for example demonstrated in the context of planet formation by pebble accretion by \citet{Brugger18}, 
who  compared the resulting mass function obtained using Eq. \ref{B15} on one side, and computed solving internal structure equations on the other side. 
These authors showed that the resulting planetary mass was very different in the two cases. When using the approximate formula (Eq. \ref{B15}), 
many planets were in the gas giant regime (Jupiter mass and beyond), whereas when solving properly the internal structure equations, the planet mass was in the Neptune regime.

In the present paper, we focus on these sub-critical planets, and we use {deep learning techniques to compute the critical core mass as well as the envelope mass as a function of the relevant parameters. Deep learning, specifically Deep Neural Networks  \citep[DNNs - see][]{Goodfellow-et-al-2016} is a sub-branch of machine learning, which has recently been applied to different problems in planetary sciences, from exoplanet search by transit or direct imaging method \citep{Pearson18,Gomez18}, to dynamical analysis \citep{Tamayo16,Smirnov17,Lam18}, to spectrum analysis \citep{Marquez18}.}
The advantage of using DNNs is two fold. First, the resulting critical core mass and envelope mass are close to the `real' one (meaning computed by solving the internal structure differential equations).
Second, the computer time required to compute these masses is orders of magnitude smaller. 

The plan of the paper is as follows. We first present the models we use in 
Sect. \ref{internal_structure}, in particular regarding the treatment of the micro-physics and boundary conditions. We then present the architecture of the DNNs we use, as well as the data we 
use to train it (Sect. \ref{DNN}). Finally, in Sect. \ref{results} we compare the results of the DNN with the ones obtained by solving the internal structure equations.
All the tools developed to use the DNNs are provided as Jupyter notebooks at \url{https://github.com/yalibert/DNN_internal_structure/} for an easy implementation in python codes.

\section{Internal structure models}
\label{internal_structure}

\subsection{Planetary growth}
In the core accretion model, a planet is formed first by the accumulation of heavy elements in a central core, followed by gas accretion, which builds an envelope on top of the core. 
The boundary between the protoplanet and the disc is in principle arbitrary, usually taken as a mean between the Bondi and the Hill radius \citep{Lissauer09}. 
Between two time steps the envelope mass grows by two mechanisms:

 (i) The increase in core mass allows to bind, in hydrostatic equilibrium, a more massive envelope around it.
 
 (ii) The envelope radiates energy into the disk, cooling, and therefore, contracting. This allows for more gas to fill the radius of the protoplanet. 
 
 The standard procedure to compute the growth of a planet is to solve the internal structure equations (Sect.\ref{equations}) from the top of the envelope to the core-envelope boundary. 
 The energy radiated away in each time step is calculated from energy conservation. This provides the updated luminosity $L$ required to solve the equations \citep[see][]{Venturini16}.

\subsection{Equations}
\label{equations}

In order to generate the data used to train and validate our DNNs, we solve first the internal structure equations \citep{Kippen13} for a large set of boundary conditions. 
For this, the protoplanet is modelled as composed by an inert core of mean density $\rho_{\rm core}$ = 5 g/cm$^3$,  surrounded by a gaseous envelope with a certain composition (see below).
For the structure equations, we use $v = r^3 / M_{\text{r}}$ as the independent variable, where $r$ is the radial coordinate of the planet, and $M_{\text{r}}$ the mass enclosed inside a sphere of radius $r$. 
This variable is convenient to stop the integration of the equations at the exact core density and avoid interpolations. Thus, the equations are written as \citep{Venturini15}:
\begin{align}\label{structure_eqs}
 \frac{dP}{dv}  = & - G \rho \bigg(\frac{M_{\text{r}}}{v^2}\bigg)^{2/3} [3 - 4 \pi \rho v]^{-1},  \\ 
 \frac{dM_{\text{r}}}{dv}  =  &  4\pi \rho M_{\text{r}} [3 - 4 \pi \rho  v]^{-1}, \nonumber \\
 \frac{dT}{dv}  = &  \frac{T}{P}\frac{dP}{dv}\nabla , \nonumber
\end{align}
where $P$ is the pressure, $\rho$ the density, $T$ the temperature, and $G$ the gravitational constant. 
Following the Schwarzschild criterion, the gradient in Eq.(\ref{structure_eqs}) is:  $\nabla = {\rm min}(\nabla_{\rm conv}, \nabla_{\rm rad})$, 
where
\begin{equation}\label{ad_grad}
 \nabla_{\text{ad}}=\bigg(\frac{\partial \text{ln} T}{\partial \text{\text{ln}} P}\bigg)_{\text{S}}
\end{equation}
is the adiabatic gradient, and the radiative gradient is given by: 
\begin{equation}\label{rad_grad}
 \nabla_{\text{rad}}= \frac{3 \kappa L P}{64 \pi \sigma G M_{\text{r}} T^4},
\end{equation}
$\sigma$ being the Stefan-Boltzmann constant, $\kappa$ the opacity and $L$ the luminosity. 

The luminosity is assumed uniform throughout the envelope, and results from the energy released by the accretion of solids and the contraction of the envelope \citep[see detailed explanation in][]{Venturini16}.

To close the system of structure equations, an equation of state (EOS) must be provided. The EOS gives $\rho$ and $\nabla_{\text{ad}}$ as a function of $T$, $P$ and composition. 

\subsection{Boundary conditions}
\label{boundaryconditions}
The outer boundary conditions to solve Eqs. \ref{structure_eqs} are set by the temperature ($T_{\rm out}$) and pressure ($P_{\rm out}$) at the planetary radius, 
defined as a combination of the Bondi and Hill radius \citep{Lissauer09}: 
\begin{equation}
	R_{\rm P} = \frac{G \mplanet}{ \left( C_{\rm S}^2 + 4 G \mplanet / \rhill \right) } 
\end{equation}
where $C_{\rm S}^2 $ is the square of the sound velocity in the protoplanetary disk at the planet's location $a$,
and $\rhill = a_{\rm p} \left( \frac{ \mplanet}{3 M_\odot } \right)^{1/3}$. 

The inner boundary conditions are given by the planetary core mass and core radius.
When solving these differential equations in planet formation models \citep[see e.g.][]{Alibert05}, the total mass of the planet is iteratively adjusted until all the boundary conditions are fulfilled.

\subsection{Microphysics}
\label{microphysics}

In this work we assume the envelope to be composed by a mixture of hydrogen and helium, with a helium mass fraction of $Y=0.28$. We do not consider the
effect of pollution by accreted solids \citep{Venturini15,Venturini17}. Such an effect will be included in a future work.
The density and adiabatic gradient of the internal structure equations are determined by $T$, $P$ and composition, through the equation of state (EOS). 
Since we assume a H-He composition for the envelope, we adopt the EOS of \citet{SCVH}.
{ For temperatures and pressures below the limits of the  \citet{SCVH} table, the EOS of ideal gas for the mixture of H$_2$ and He is assumed, accounting for the corresponding degrees of freedom for the entropy and energy of the mixture.}

The opacities are defined as:
\begin{equation}\label{kappa}
\kappa = \max \, (f_{\rm dust} \, \kappa_{\rm dust}, \kappa_{\rm gas})
\end{equation}
where the dust opacities are taken from \citet{BL94} (hereafter BL94), and  the gas opacities from \citet{Freedman14}. 
The factor $f_{\rm dust}$ accounts for the possibility of the dust opacities to have reduced values compared to the ISM. This can result due to grain growth and settling \citep{Movshovitz10, Mordasini14}.
In this work we present calculations for $f_{\rm dust}=1$ and $f_{\rm dust}=0.01$.
Since the dust opacity values peak at the top of the envelope and the gas ones at the deeper, hotter regions, the definition in Eq.(\ref{kappa}) is equivalent to considering $\kappa = f_{\rm dust} \, \kappa_{\rm dust} + \kappa_{\rm gas}$, as shown in \citet{Mordasini14}. 

The domain of validity of the Freedman gas opacities in pressure and temperature is, respectively, 1 to 300 bars and 75 K to 4000 K.  
At the core-envelope boundary, higher values of temperature and pressure typically exist, but in that domain of high temperature and pressure the
envelopes are always convective. Therefore, even taking a constant extrapolation in these region does not affect the internal structure computation. 
An analytical fit to the Freedman gas opacities is given in \citet{Freedman14}. This is implemented to reduce computational time.

\section{Deep Neural Network}
\label{DNN}

\subsection{Principle}
\label{principle_DNN}

In the last decades, \textit{machine learning} has seen a tremendous development, thanks to the advancement of novel techniques, and to the availability of very large amount of data. 
In particular, \textit{neural networks} have been more and more used during the last decade to achieve very good performances in tasks such as image classification and text translation.
This method has also been applied to \textit{regression}, which is linked to the prediction of a continuous variable \citep[see][]{Goodfellow-et-al-2016}, and is the case of interest in this paper. In regression,  
the task of a neural network is to use some input variables (like temperature, pressure at the planetary outer radius, etc.) to predict one { or more} variables (the mass of the gas envelope in our case).
In the machine learning literature, the input variables are referred as \textit{features} and the predicted variable as \textit{target variable}.

A neural network is made of different layers of units, each unit performing a simple task: linear combination of its inputs, followed by a non-linearity (see below). 
Each layer of a neural network takes typically as input the outputs of the previous layer to produce a series of outputs that will be processed by the next layer. 
The first layer takes as input the features, and the last layer outputs a prediction of the target variable that is, hopefully, close to  the actual target variable. A 
neural network is said to be \textit{deep} (named DNN for Deep Neural Network) when it has at least one hidden layer (i.e., a layer that is not the first or the last one). A DNN 
is trained by providing a (large) number of data (features and target variable), and adjusting the coefficients (also named weights)  of all linear combinations 
of all units  in an iterative way, in order to achieve the best match between the predicted target value and the actual one.
{The matching between the predicted and actual values is quantified using a cost function (see below), which is a function of all the parameters of the DNN. The gradient of the cost function
is then computed (as a function of the parameters of the DNN) using a method called backpropagation \citep{Goodfellow-et-al-2016}, which consists basically in the application of the chain rule for derivation. Finally, gradient descent is used in order to find the optimum parameters. Note that the cost function is in general a non-convex function of the DNN parameters, and many local minima can exist. There is therefore no guarantee that the parameters found are the ones leading to the absolute best performances of the DNN.}

We have considered two DNNs. The first one, DNN1, to predict the critical core mass (as a function of four parameters, see next section). The second one, DNN2, to predict the gas 
envelope mass, as a function of five parameters (the same four parameters plus the core mass). The architecture of the DNNs for the different values of $f_{\rm dust}$ is the same, but the resulting
weights are of course different.

DNN1 has 3 hidden layers,  each having 128 units. These numbers were found by a series of trial and error tests {where we varied the number of layers and the number of units per layer in a grid search (see Sect.\ref{comparison_other_methods}). We also compare in  Sect. \ref{comparison_other_methods} the results obtained using other machine learning methods.}
The DNN is  fully connected, 
meaning that each unit of a layer is connected to each layer of the previous and next layer. For each unit, we chose to use 
the ReLU function for the non-linearity, given by:
\begin{equation}
 ReLU(x) = \max (x,0)
\end{equation}
The cost function we minimise is the mean square error between the predicted and the actual critical mass:
\begin{equation}
C(w) = {1 \over N} \sum\limits_{k=1}^{N} \left( M_{\rm critical, predicted, k} - M_{\rm critical, target, k} \right)^2
\end{equation}
where $N$ is the total number of data points we use, $M_{\rm critical, target, k}$ is the critical core mass computed using the internal structure differential equations
(see previous section),  $M_{\rm critical, predicted, k} $ is the value predicted by the DNN, and $w$ are the weights. 

DNN2 has five hidden layers, each having 128 units, the non-linearity is also the $ReLU$ function, and the cost function is similar, except that we compare the 
predicted envelope mass and the real one. 

Both DNN were trained using the ADAM optimiser \citep{geron} for 10000 training epochs. {The learning rate is chosen equal to 0.0001 for the training. Larger learning rate could be chosen, but this would translate to noisy decrease of the cost function. The other parameters of the ADAM optimiser are taken as the default values in TensorFlow (see \url{https://www.tensorflow.org/api_docs/python/tf/train/AdamOptimizer}).} We did not used mini-batch since we could fit the whole training dataset (of size 25000 and 800000 respectively for DNN1 and DNN2) in the memory at once. The sizes of the test and validation sets are both equal to 1000 for DNN1 and 300000 for DNN2. {Finally, the bias for all units was initialised at 0, whereas the weights were randomly initialised using a gaussian distribution centered on 0, with a variance scaling with the square root of the number of input and output connections \citep{geron}.}

\subsection{Data}
\label{data}

To generate the data, we selected, in a first step,  $\sim$10000 points in a four dimension space, by drawing at random a semi-major axis $a$ (uniform in log between  0.1 and 30 AU), a pressure $P_{\rm out}$
(uniform in log between  $10^{-2}$ and $1500$ dyn/cm$^2$.), a temperature $T_{\rm out}$ (uniform between $30$ K and $1500$ K),  and a luminosity $L$ (uniform in log between $10^{22}$ erg/s and $10^{28}$ erg/s).
These luminosities encompass the energy per unit time that would be given by an accretion rate of solids of  $\Mzdot = 10^{-10} \, \mearth$/yr onto a core of 1 $\mearth$, and 
$\Mzdot = 10^{-5} \, \mearth$ /yr onto a core of 20 $\mearth$. 

For each set of $(a,L,P_{\rm out},T_{\rm out})$, we compute, by solving the differential equations presented in Sect. \ref{internal_structure}, the core mass for a given planetary mass. 
$P_{\rm out}$ and $T_{\rm out}$ are used as the outer boundary conditions (Sect. \ref{boundaryconditions}).
For a planetary mass growing from $M_{\rm planet}$ = 1 $\mearth$ onwards, the core mass first increases, then reaches a local maximum and
then decreases. The \textit{critical core mass} is simply the first local maximum core mass obtained in such a way \citep{mizuno80,PT99}.

For each of these $(a,L,P_{\rm out},T_{\rm out})$, we compute a full $M_{\rm planet}-M_{\rm core}$ curve, for a planetary mass ranging from 1 $\mearth$ up to the corresponding critical mass.
Each of these curves is sampled with a step of 0.01 $\mearth$ in $M_{\rm planet}$, and the resulting $M_{\rm envelope}$ as a function of the five parameters $(a,L,P_{\rm out},T_{\rm out},M_{\rm core}$)
is finally computed. In addition to the relation between the envelope mass and the five input parameters $(a,L,P_{\rm out},T_{\rm out},M_{\rm core})$, these curves provide us with the critical core
mass as a function of the four  first parameters $(a,L,P_{\rm out},T_{\rm out})$.

Importantly, since the critical core mass depends on the $(a,L,P_{\rm out},T_{\rm out})$ parameters (it is e.g., smaller for smaller $L$), the number of points on a $M_{\rm planet}-M_{\rm core}$ is 
not the same for each $(a,L,P_{\rm out},T_{\rm out})$ choice.  Since the cost function $C(w)$ defined above for DNN2 gives the same importance to each data points, taking all the points of
 all the $M_{\rm planet}-M_{\rm core}$ curves would result in increased performances for large  critical core masses than for small ones. Indeed, data points for which $M_{\rm critical} (a,L,P,_{\rm out}T_{\rm out})$ 
is large have more importance in the cost function, since they are more numerous. Such an approach could lead to a DNN  predicting accurately the envelope masses for, say, large $L$ (large critical cores) and 
 very poorly for low $L$ (small critical cores). To avoid this effect, and have a similar prediction accuracy for all parts of the  $(a,L,P_{\rm out},T_{\rm out},M_{\rm core})$ space, we use the same 
 number of points for each  of the  $M_{\rm planet}-M_{\rm core}$ curves (this technique is known as \textit{stratification}). Finally, before implemented, all the data are pre-processed by removing the mean and 
 scaling to unit variance each in each dimension separately.

\begin{figure*}
\centering
\includegraphics[width=0.99\textwidth]{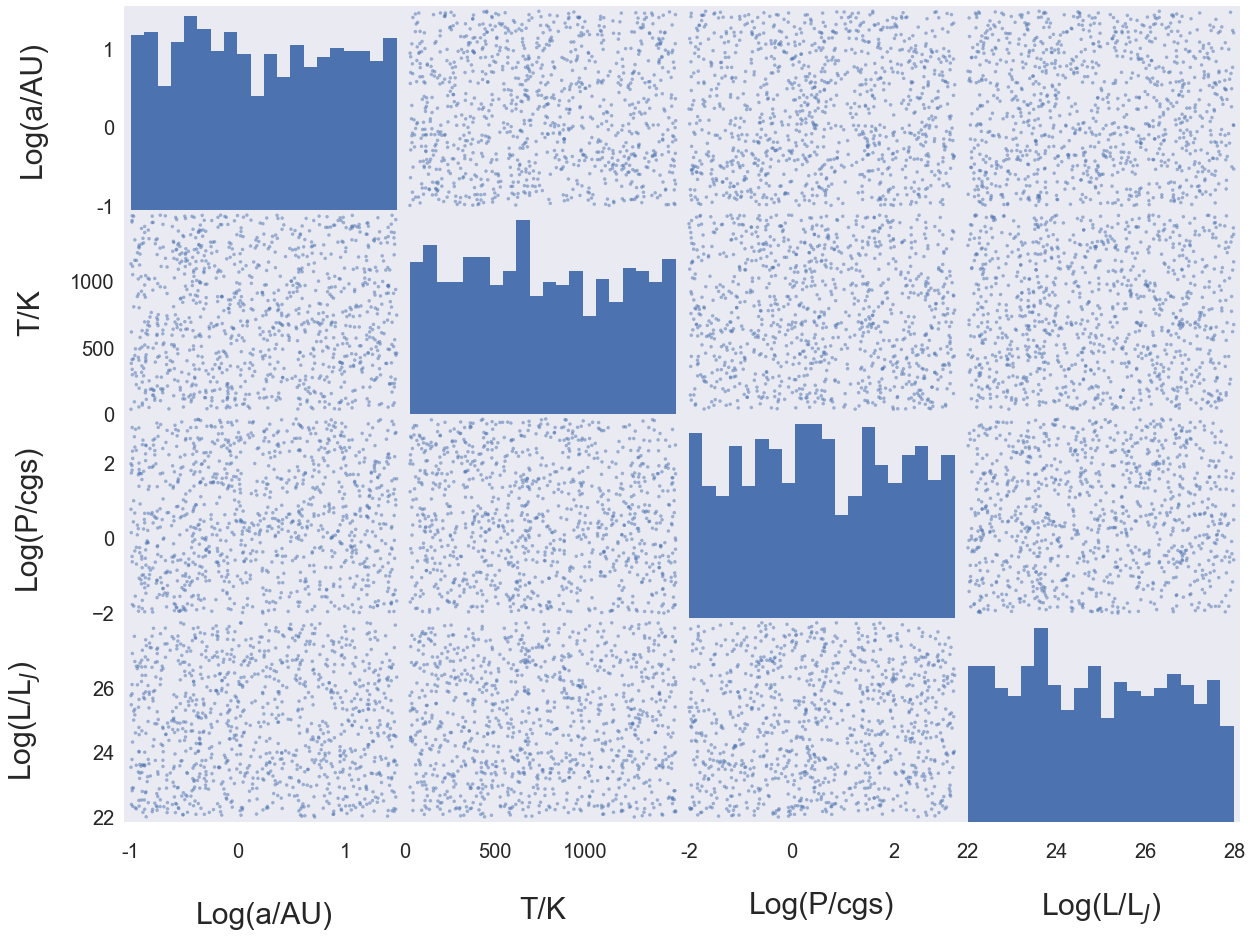} 
\caption{Initial conditions for a subset of 1000 datapoints. Each panel shows the correlation between two quantities, the panels on the diagonal showing the histograms of the different quantities.}
  \label{IC_all}
\end{figure*}

We obtain $\sim$27000 data points to be used with DNN1 (for predicting the critical core) for each choice of the opacity parameter $f_{\rm dust}$, 
and $\sim10$ millions data points to be used to train DNN2 (to predict the envelope mass), again for each choice of the opacity parameter $f_{\rm dust}$. 
{ Fig. \ref{IC_all} shows the distribution of $(a,L,P_{\rm out},T_{\rm out})$ for a subset of the datapoints.}

\begin{figure*}
\centering
\includegraphics[width=0.99\textwidth]{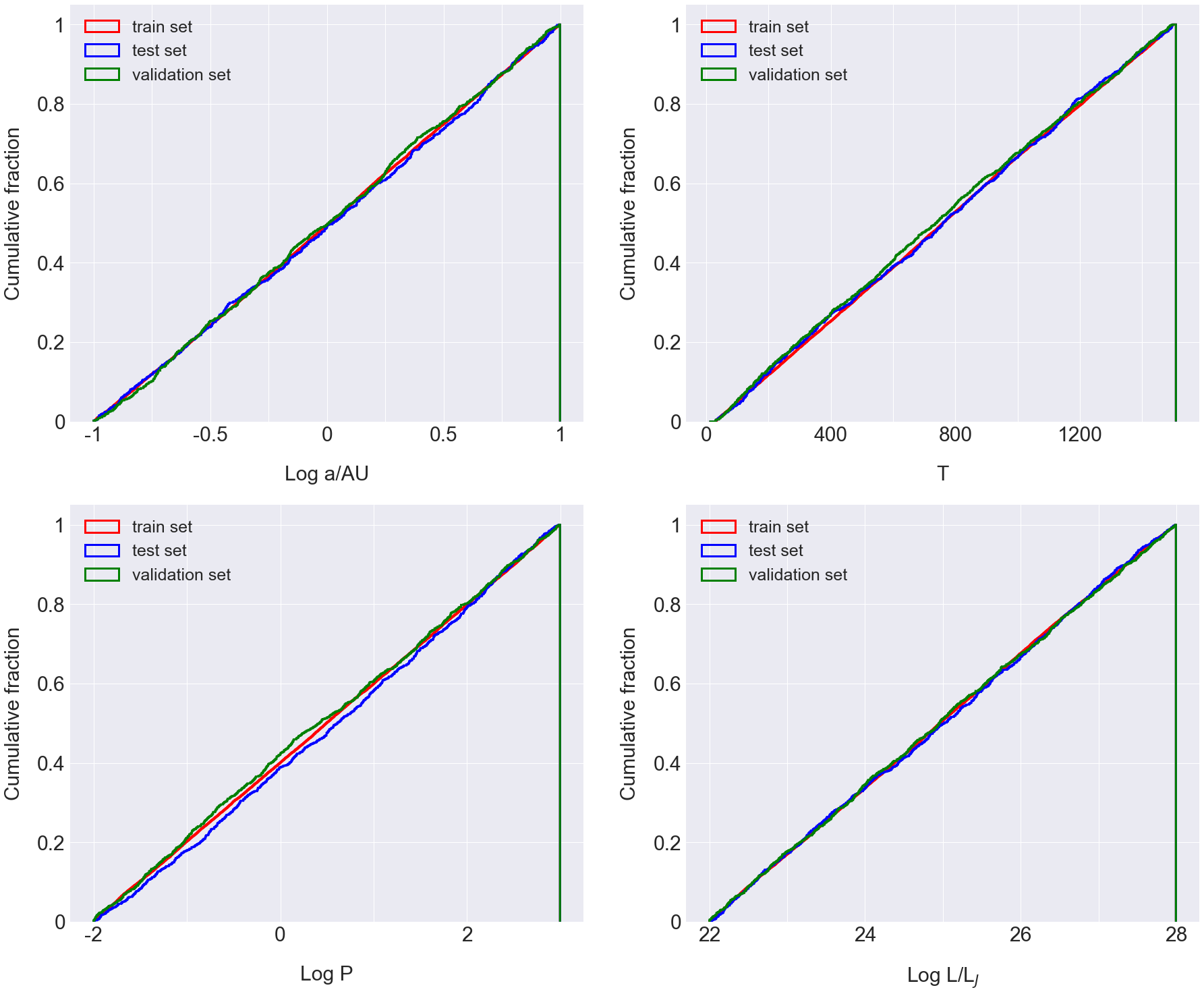} 
\caption{Cumulative distribution of the input parameters for DNN1, for the training, validation and test sets.}
  \label{comparing_distributions}
\end{figure*}

These data points are split randomly in a training set, a validation set and a test set. The training set is used to adjust the weights of the DNNs, the validation set is used 
to estimate the generalisation performances of the DNNs, and avoid, in particular, overfitting. Once the optimum weights of the DNNs are computed (in practical when the cost function on the 
validation set is the smallest), we use the test set to present the results shown in the rest of the paper (the data points used to generate the different plots shown below are 
therefore never used in any part of the optimisation process of the DNNs). {It is important to check that the distribution of the parameters is the same for the three sets (training, test and validation). This should be the case, as the sets were generated randomly, and this can be checked looking at the cumulative distribution of the input parameters for the three sets as shown in Fig. \ref{comparing_distributions}.}

{The $(a,L,P_{\rm out},T_{\rm out},M_{\rm core})$ parameters are drawn at random independently in our data generation procedure. It is clear, on the other hand, that some of these parameters are strongly correlated, for exemple the pressure and the temperature are a function of the semi-major axis in actual proto-planetary discs. This raises two questions: 1- what kind of internal structure could unrealistic input parameters (e.g. high temperature and low pressure) lead to, and 2- what is the effect of the these unrealistic data on the global optimisation process of the DNN. Indeed, the cost function as presented above takes into account all considered data with an equal importance. Taking into account unrealistic data could therefore degrade the performances of the whole DNN.}

\begin{figure}
\includegraphics[width=0.45\textwidth]{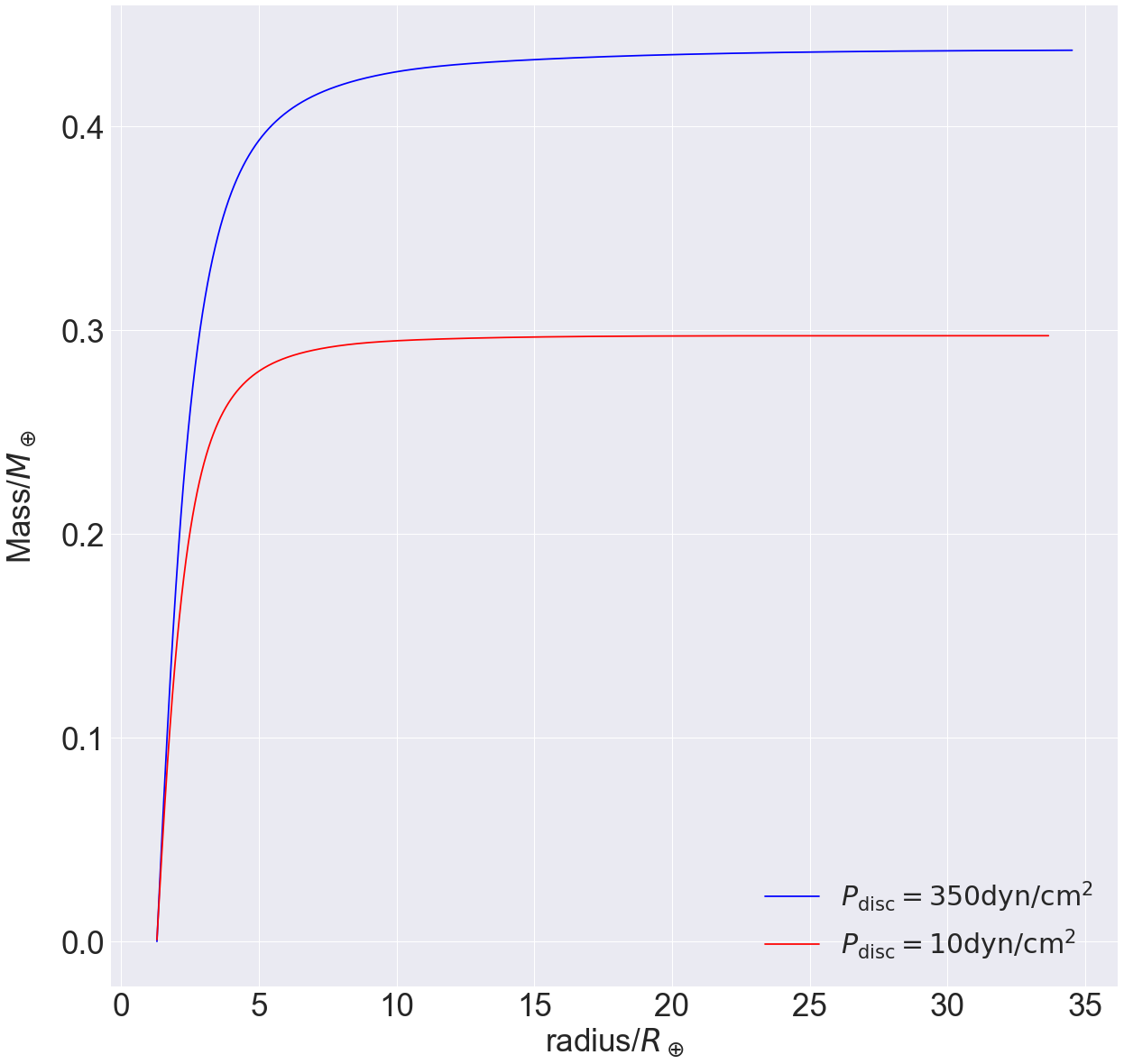} 
\caption{Envelope mass as a function of the radius for two models. For both models, a planetary core of mass equal to 2 $\mearth$ is located at 0.5 AU, with a luminosity equal to $10^{25}$ erg/s and a temperature at the planet-disc interface equal to 400K. The pressure at the disc-planet interface is equal to 10 dyn/cm$^2$ (an unrealistic value) or 350 dyn/cm$^2$.}
  \label{exemples_IS}
\end{figure}

{Regarding the first question, we have checked that our internal structure code converges even for unrealistic cases. As an example, Fig. \ref{exemples_IS} shows the internal structure (envelope mass as a function of the radius) for two cases. The planetary core mass is equal to 2 $\mearth$, the semi-major axis equal to 0.5 AU, and the luminosity equal to $10^{25}$ erg/s, which corresponds to an accretion rate of $\sim  5.5 \times 10^{-8} \mearth/$yr . The temperature is equal to 400K and the pressure is equal to 10 dyn/cm$^2$ in the first (unrealistic) case, and equal to 350 dyn/cm$^2$, in the second (more realistic) case.  These values are typical for a protoplanetary disc \citep[see][]{Venturini17}. The envelope mass in the unrealistic case is equal to 0.261 $\mearth$, to be compared to 0.437 $\mearth$ in the realistic case. }

{We address the second question (the potential effect of unrealistic data on the DNN performances) in Sect. \ref{comparison_other_methods} below.}

\section{Results}
\label{results}

\subsection{Critical mass}
\begin{figure*}
\centering
\includegraphics[width= 0.99\textwidth]{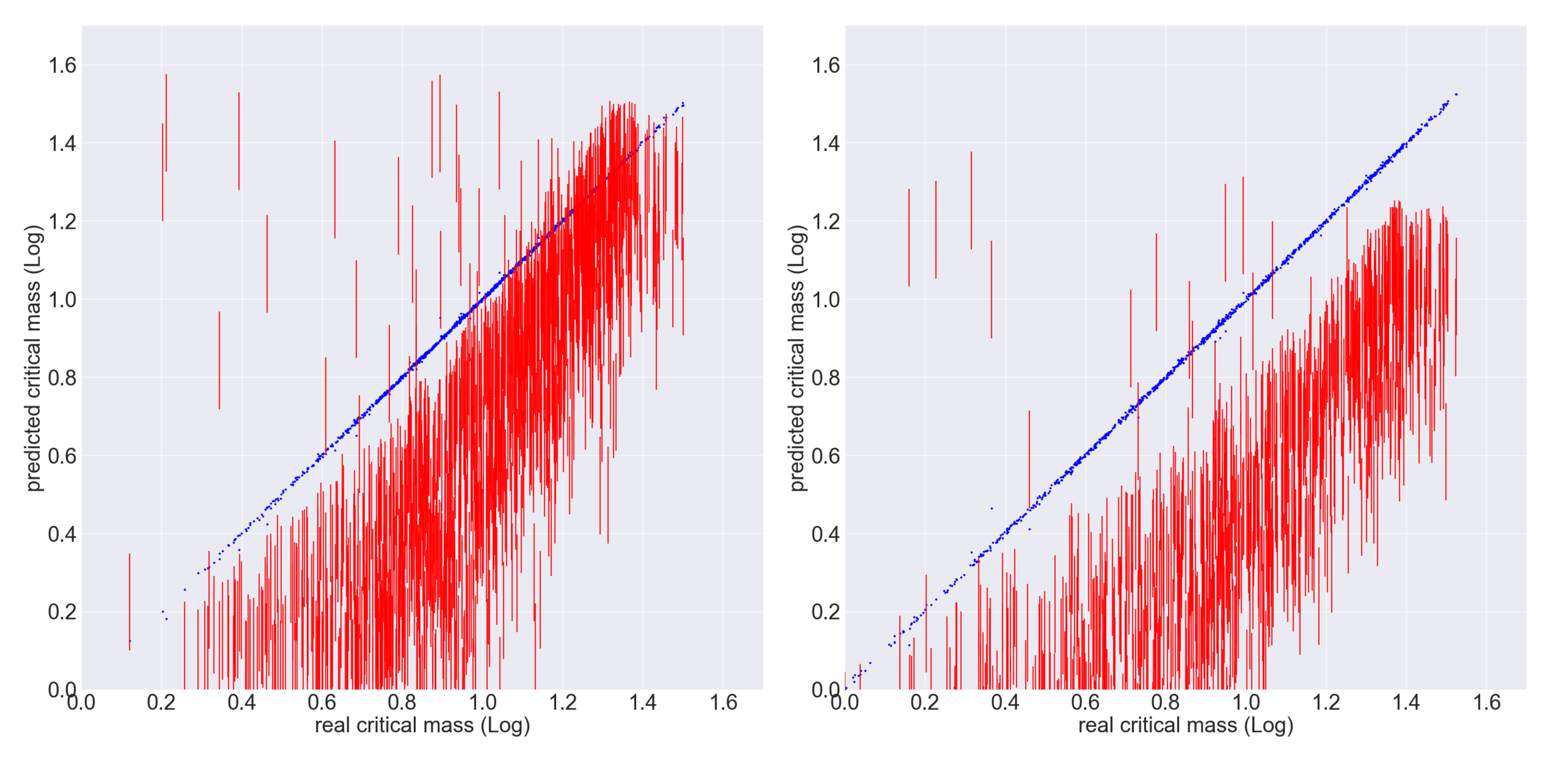} 
\caption{Predicted critical mass (in Earth masses, logarithmic scale) as a function of the critical mass obtained by solving the internal structure equations (blue),
and using the IL04 formula (red - see Eq. \ref{Mcrit_IL04}, with  exponents of the formulas  equal to 0.25).  \textit{Left:} using the ISM opacity from BL94. The lower end of each line corresponds to the result obtained for an
opacity of 1 cm$^2$/g, the higher end for an opacity of 10 cm$^2$/g. \textit{Right:} reducing the BL94 opacity by a factor 100. The lower end of each line corresponds to the result obtained for an
opacity of 0.01 cm$^2$/g, the higher end for an opacity of 0.1 cm$^2$/g.}
\label{Mcrit_DNN_IL04_Z0}
\end{figure*}


The DNNs are trained using the Tensorflow library (\url{https://www.tensorflow.org}), the notebooks to use the DNN are available at \url{https://github.com/yalibert/DNN_internal_structure/}.

As mentioned above, the DNNs are trained using only a sub-sample of the generated data, the other part is split between a test set, and a validation set. We use the test set to compute
the difference between the predicted critical core mass and the one derived using internal structure equations. The correlation between the two masses is shown in Fig. 
\ref{Mcrit_DNN_IL04_Z0}, our results are plotted in blue whereas the critical mass from IL04 is shown in red (Eq. \ref{Mcrit_IL04}).
We emphasise the fact that, in the resolution of the internal structure equations, the envelope opacity is not constant but rather a function of the local temperature and 
pressure, and varies between $\sim \!\!  1$ and $\sim \!\!  10$ cm$^2$/g in the radiative layers of the atmosphere ($T\lesssim1000$ K). 
Hence, we use these two limiting values to compare our results with the ones of IL04, 
and plot the resulting critical mass as a vertical line in Fig. \ref{Mcrit_DNN_IL04_Z0}.
As can be observed, the critical mass computed using Eq. \ref{Mcrit_IL04} has a much larger variance compared to the results obtained using the DNN, and is general under-estimated.
Using our DNN, the average difference between the predicted critical core mass and the one obtained by solving the internal structure equations is $\sim \!\!  1.67 \%$, with a majority ($99.3 \%$) 
of the critical masses being predicted with an accuracy better than 5\%, and a maximum error of $\sim \!\! 14.2 \%$ for critical masses larger than 1 $\mearth$.
 
We perform the same comparison on the right panel of Fig. \ref{Mcrit_DNN_IL04_Z0},  now assuming that the opacity is reduced by a factor 100 compared to the one of BL94. 
When using the IL04 formula, the opacity ranges from 0.01 to 0.1 cm$^2$/g. In this case, the critical core mass obtained using Eq. \ref{Mcrit_IL04} is substantially biased towards smaller values.

\subsection{Envelope mass}
\begin{figure*}
 \center
\includegraphics[width=0.99\textwidth]{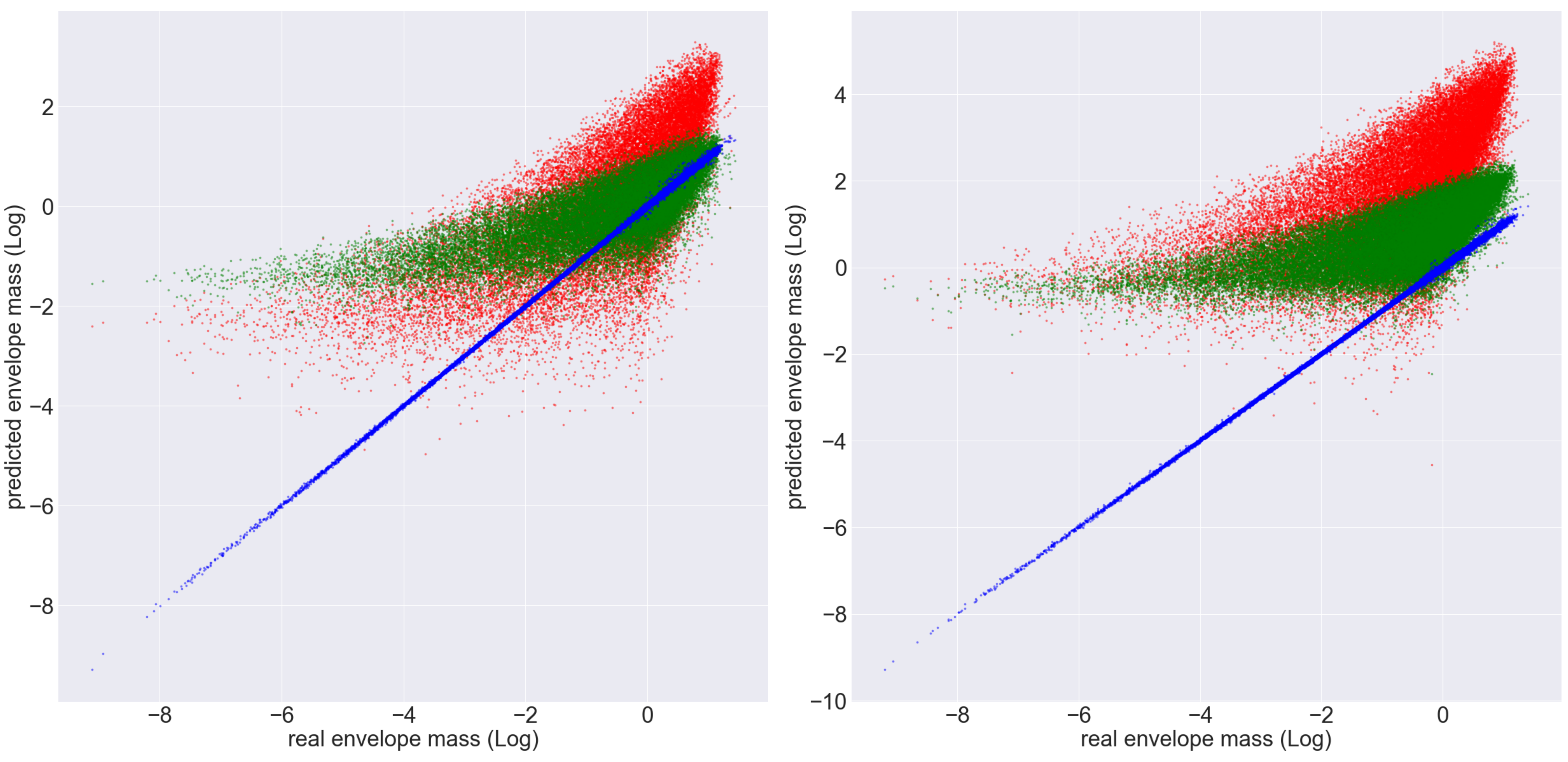} 
\caption{Predicted envelope mass (in Earth masses, logarithmic scale) as a function of the envelope mass obtained by solving the internal structure equations (blue). 
The green and red points show the estimation of the envelope mass using the I00 and B15 accretion rate, respectively. This accretion rate is 
integrated on the disc lifetime, itself assumed to be randomly uniformly distributed between 0 and 5 Myr. Only envelope masses for core masses larger than 1 $\mearth$ are represented. 
\textit{Left:} using the ISM opacity from BL94. The envelope mass using I00 and B15 is computed assuming an opacity of 1 cm$^2$/g. \textit{Right:} reducing this opacity by a factor 100. The envelope mass using I00 and B15 is computed assuming an opacity of 0.01 cm$^2$/g.}
  \label{Menve_DNN_PY14}
\end{figure*}

Similar to the critical mass prediction, we use the test set of envelope masses to compute the difference between the predicted envelope mass and the one derived using internal structure equations. 
The correlation between the two masses is shown in Fig. \ref{Menve_DNN_PY14}, where we also show the result estimated using Eqs. \ref{TKH} (I00) and \ref{B15} (B15). 
In these two latter cases, we need to specify the timescale over which gas is accreted, as these two formulas only provide a gas accretion rate. 
For the simple estimates provided in Fig. \ref{Menve_DNN_PY14}, we have  assumed that gas is accreted on a timescale that is uniformly distributed between 0 and 5 Myr.  
This timescale should be equal to the disk lifetime. We emphasis the fact that, in a real situation, the core mass of the planet could increase during this timescale, therefore 
modifying the accretion rate of gas. In the estimates provided in the figures, this effect is not included.

Comparing the three different cases, we see that the variance is much larger when using the B15 and I00 formulas. Some of this variance of course results from the assumption 
of the distribution of the time during which gas is accreted. We remark that, as for the case of the critical mass, it is not obvious which value of the envelope opacity should 
be used in order to obtain the most meaningful comparison between the DNN and the fitting formulas of I00 and B15. 

Using our DNN, the masses are predicted with an accuracy better 
than 20\% in more than $97 \%$ of the cases.

\subsection{Planet formation tracks}
\begin{figure*}
\centering
\includegraphics[width=0.99\textwidth]{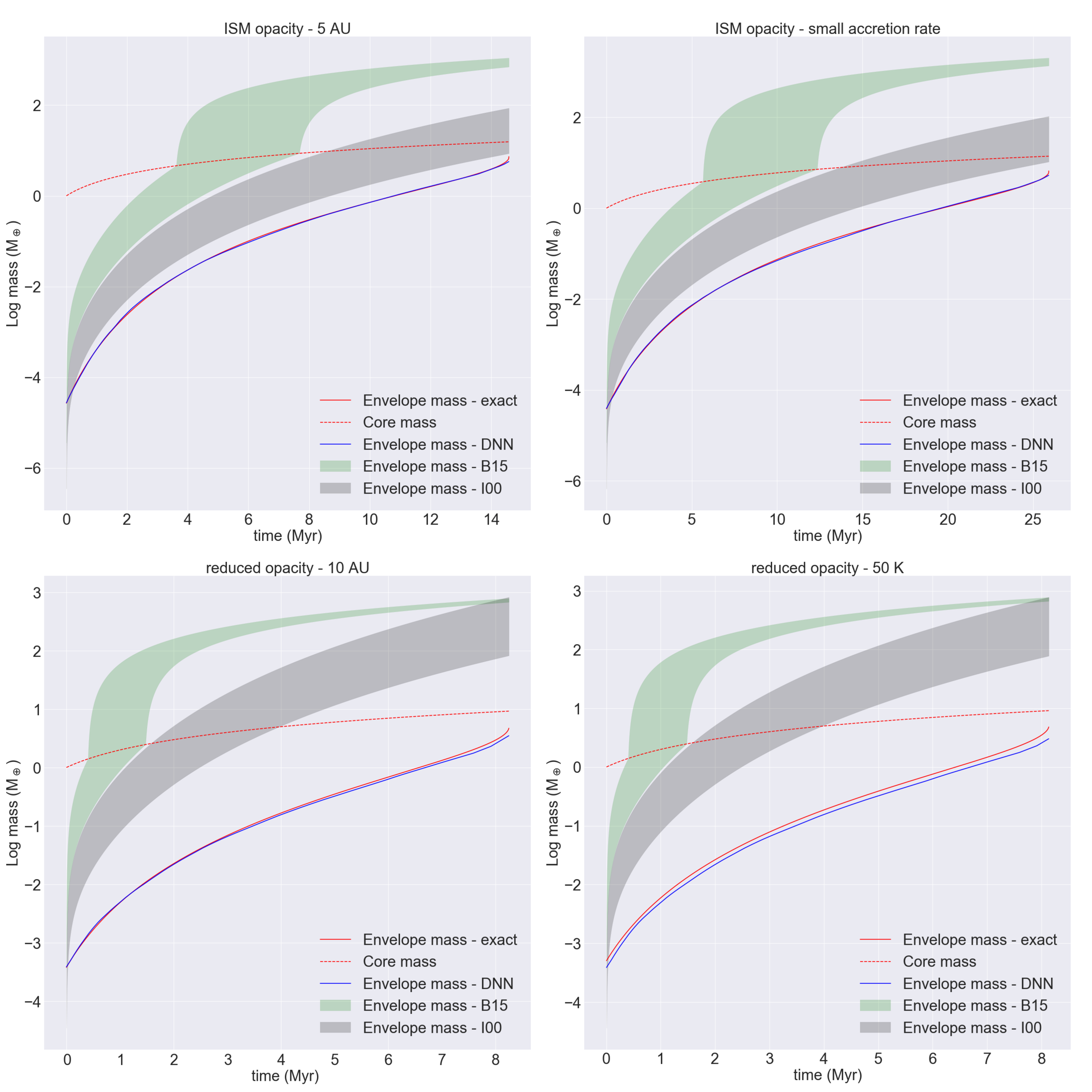} 
\caption{Evolution of the envelope mass as computed solving the internal structure equations (red lines), and using our DNN (blue lines). We also show the comparison with what would be 
obtained using the approach of I00 (gray areas), and B15 (green areas) for different choices of the envelope opacity (see text). The evolution of the core mass is shown by the red dashed line. 
Note that each figure has different scales for the time and the envelope mass. \textit{Upper left}: using BL94 opacity for a planet at 5 AU accreting with $\dot{M}_{\rm core} = 10^{-6} \mearth/$ yr.
\textit{Upper right}: using BL94 opacity for a planet at 5 AU accretion with $\dot{M}_{\rm core} = 5 \times 10^{-7} \mearth/$ yr. \textit{Lower left}: using reduced opacity for a planet at 10 AU accreting with $\dot{M}_{\rm core} = 10^{-6} \mearth/$ yr. \textit{Lower right}: using reduced opacity for a planet at 5 AU accreting with $\dot{M}_{\rm core} = 10^{-6} \mearth/$ yr, but assuming a disk temperature equal to 50 K (150 K in all other cases). The kink in
the green areas is due to the rapid runaway accretion of gas when the planet reaches the cross-over mass (mass of accreted gas equal to mass of accreted solids - see B15).}
  \label{tracks}
\end{figure*}
In order to compare in a more realistic way the result of the DNN and the one coming from the resolution of internal structure equations, we present in Fig. \ref{tracks} 
planetary growth tracks for different cases. We consider different formation locations (5 and 10 AU), different accretion rates ($\dot{M}_{\rm core} = 10^{-6} \mearth/$ yr and $\dot{M}_{\rm core} = 5 \times 10^{-7} \mearth/$), 
different temperatures (50 and 150 K) and two dust opacities. For each case, we also compare the evolution of the envelope mass with the one that would be obtained with the approach 
of I00 (Eq. \ref{TKH}) and B15 (Eq. \ref{B15}). For these two latter cases, we consider in each case two possible values of the envelope opacity, namely $1$ and $10$ g/cm$^2$ when comparing with 
the BL94 opacity, and $0.01$ and $0.1$ g/cm$^2$ for the case where the BL94 opacity is reduced by a factor 100.

In all the cases, it is clear that the results of the DNNs are much closer to the results obtained by solving the internal structure equations, than by implementing the I00 and B15 formulas. It is also notable that the envelope mass 
is extremely over-estimated by Eq. \ref{B15}. In all the cases presented here, the planet ends up as a Neptune mass planet when using the DNN (and therefore also when solving the internal structure equations), whereas 
it ends-up in the Jupiter mass range using Eq. \ref{B15}. This result is compatible with the findings of \citet{Brugger18}. The results obtained by the formulas of I00 in the case of BL94 opacity are similar but still
somewhat larger than the one obtained from the internal structure equations. In the case of the reduced opacity, the corresponding results (using Eq. \ref{TKH}) are rather different from what is obtained by solving internal structure equations.
 We finally note that the difference between the DNNs and the internal structure equations seems 
somewhat larger in the case of the low temperature. This could be solved by using DNNs with larger layers and/or number of units per layer. The difference is however much smaller than the one obtained using either I00 or B15.

\section{Discussion}
\label{comparison_other_methods}

\subsection{Overfitting}

{
The very high accuracy that we obtain may seem surprising, and it is important to check that the network is not overfitting the training data. In order to check this, we plot in Fig. \ref{accuracy} the training and validation accuracy as a function of the training epoch for the two DNNs that we have considered. As can be seen, the validation accuracy decreases and starts to plateau after 8000-10000 epochs. On the contrary, the training accuracy still decrease, indicating that the network start to overfit the data at this point. For that reason, the results presented in this paper are the ones obtained after 10000 epochs. {This number is high, but results from the choice of a small learning rate in the ADAM optimiser.}

\begin{figure}
 \center
\includegraphics[width=0.45\textwidth]{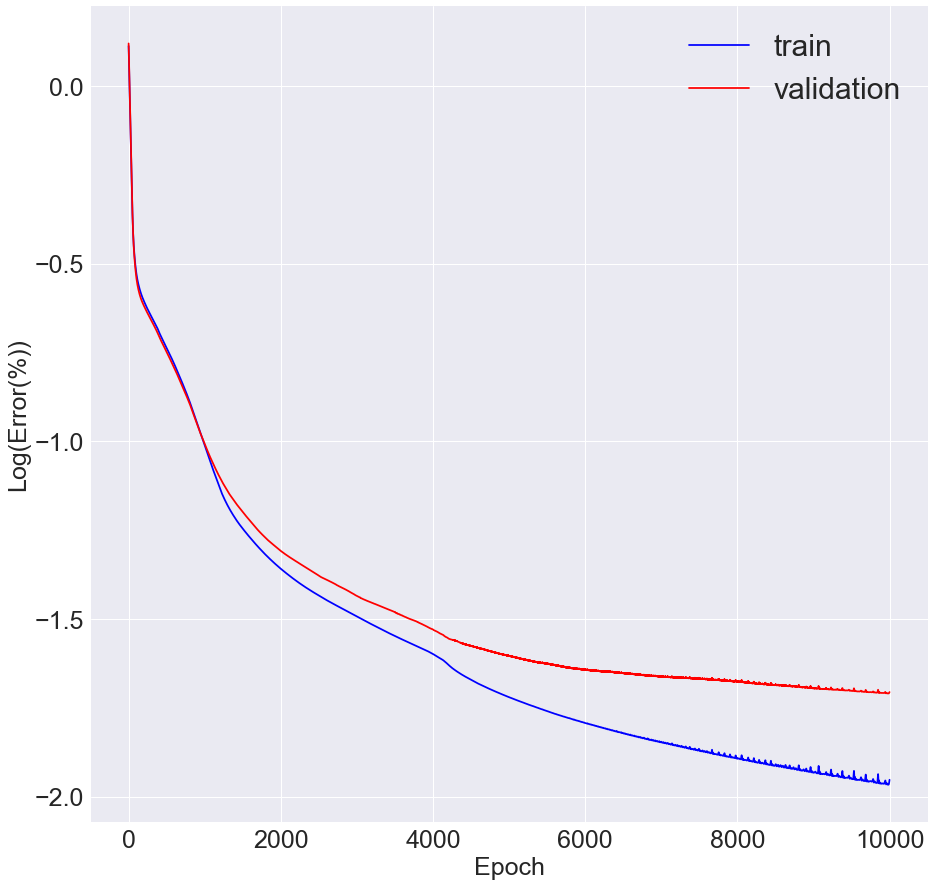} 
\caption{Training and validation accuracy as a function of the epoch for DNN1 (computation of the critical mass). }
  \label{accuracy}
\end{figure}

}

\subsection{Effect of the dataset}
{

As mentioned above, our input parameters are generated randomly in an independent way. In reality, one expects that, for exemple, the temperature and pressure are strongly correlated with the semi-major axis. As discussed previously, including such unrealistic data-points could degrade the performances of the DNN. In order to test this, we have considered two cases where we trained DNN1 over 10000 epochs, with the same initialisation of the weights. In the first case, we have considered all the data we computed (including the unrealistic ones), whereas in the second case, we have only considered data-points fulfilling the two following conditions:
\begin{equation}
28 \left( a \over AU \right)^{-1/2} {\rm K} < T < 2800 \left( a \over AU \right)^{-1/2} {\rm K}
\end{equation}
and
\begin{equation}
3.65 \left( a \over AU \right)^{-3.25} {\rm dyn/cm}^2 < P < 365 \left( a \over AU \right)^{-3.25}  {\rm dyn/cm}^2 
\end{equation}
These two conditions were chosen allowing for a one order of magnitude variation of the pressure and temperature below and above the values used in \cite{Venturini17}. 

We then compared the output of the two trained DNN on the same set of validation data-points (304 points). In order to have a fair comparison, all validation data-points fulfil the above conditions and are therefore `realistic', and both DNNs were trained using the same number of training data-points (5738). We obtained an accuracy of 4.86 \% in the case of the full dataset, and 4.38 \% in the case of the reduced dataset. Using only 'realistic' data-points indeed improves the performances of the DNN as expected, although the change is not massive. The accuracies reached in these two cases are not as good as in the previous section, since both DNNs are trained using a reduced number of points, which degrades their performances.
}

\subsection{Comparison with other machine learning methods.}
{

We  compared our results with the ones obtained using other machine learning methods:
\begin{itemize}
\item linear regression
\item random forest with 10,100, 1000 and 10000 trees
\item gradient boosting (using the XGBoost library - see \url{https://xgboost.readthedocs.io/en/latest/})
\item Support Vector Regression
\item different architectures of the DNN
\end{itemize}
For each of these methods, we used the logarithm of the input quantities (except the temperature), and the temperature, as features.

All these methods are described in the literature \citep[e.g.][]{geron}. For each of these methods, a number a hyper-parameters can be tuned in order to improve the performances. For all the methods we  considered, we  kept the default parameters as defined in scikit-learn (see \url{https://scikit-learn.org/stable/}), except when explicitly mentioned. An exhaustive evaluation of the influence of all these hyper-parameters is out of the scope of this paper.

{ We have in addition, in the case of gradient boosting, performed a random grid search in order to infer if some hyper-parameters could lead to better results. We have ran 100 models, varying the boosting method (`gbtree', `gblinear' or `dart', see \url{https://xgboost.readthedocs.io/en/latest/index.html}), the maximum depth of the trees (from 1 to 8), the learning rate (0.05, 0.1 or 0.2), the number of estimators (30, 100 or 300). For each of these 100 cases, we have used a 5-fold cross-validation. The best model has been found to correspond to the dart boosting method, a learning rate of 0.05, a maximum depth of trees of 6, and 300 estimators. In this case, the error is of 6.82 \%. This is in any case around 3 times worse than the best result we obtained with our baseline DNN.}

In all the cases, we  used the same data-points for training and testing (including some 'unrealistic' datapoints, see above), and we compared the resulting accuracy to the one obtained with our baseline DNN1 model (3 hidden layers of 128 units each). The results are presented in Table \ref{results_ML_crit}. As can be seen in the table, the DNN outperforms all other methods we  considered by at least a factor 2 in accuracy. Interestingly enough, the DNN with  3 and 2 hidden layers gives similar results, { with however some slightly better results with 3 hidden layers}. 

 It should be noted, however, that the computer time required to train the DNN is much larger than for the other methods. For inference (computing the target value given a set of input values), the required computer power using DNN is also larger than for other methods. It remains, however, orders of magnitude smaller than the time required to solve differential equations.

\begin{center}
\begin{table}[ht]
\caption{Validation set accuracy for different machine learning methods to predict the critical mass. All DNN lines refer to fully connected Deep Neural Network, the numbers giving, respectively, the number of hidden layers, and the number of units per layer.}
\begin{center}
\begin{tabular}{lc}
\hline\noalign{\smallskip}
 baseline DNN ($3 \times 128$ )&  1.67  \% \\
 linear regression & 30.5 \% \\
  Random Forest (10 trees) & 10.5 \% \\
  Random Forest (100 trees) & 8.61 \% \\
    Random Forest (1000 trees) &  8.18 \% \\
  Random Forest (10000 trees) & 8.11 \% \\
Support Vector Regression & 12.5 \% \\
  Gradient Boosting & 13.5 \% \\
   best Gradient Boosting after random search & 6.82 \% \\
 DNN $2 \times 128$ & 1.87 \% \\
  DNN $1 \times 128$ & 13.1  \% \\
  DNN $3 \times 64$ &  2.31 \% \\
  DNN $3 \times 32$ &  3.49 \% \\
 \noalign{\smallskip}
\hline
\end{tabular}
\end{center}
\label{results_ML_crit}
\end{table}
\end{center}

}

\section{Conclusion}

We trained Deep Neural Networks to compute the critical core mass and envelope masses of forming planets, for a variety of conditions (formation location, temperature and pressure in the disc, core mass, solid accretion rate).
The resulting DNNs, which can be easily implemented with the tools we provide on github (\url{https://github.com/yalibert/DNN_internal_structure/}), give very similar results to the ones obtained by solving the internal
 structure equations, using a much reduced computer time. We also showed that some fitting formulas found in the literature in general over-estimate the resulting planetary envelope mass modestly (e.g. I00 in the case of large opacity) or very largely (e.g. using the B15 approach). Using the aforementioned formulas can therefore completely overestimate the resulting mass of a forming planet, a caveat that can be avoided using the Deep Neural Networks we provide here. 

\acknowledgements

This work was supported in part by the European Research Council under grant 239605. This work has been carried out within the frame of the National Centre for Competence in Research PlanetS supported by the Swiss National Science Foundation. The authors acknowledge the financial support of the SNSF. The plots shown in this paper made use of the seaborn software package \url{https://seaborn.pydata.org}.

\bibliographystyle{aa}
\bibliography{lit_updated2018}

\end{document}

%% file: def.tex
\def\Mzdot{\dot M_Z}
\def\text{\rm}

\def\mearth{\rm{M}_{\oplus}}

\def\f1{f_{\rm I}}

\def\beq{\begin{equation}}
\def\eeq{\end{equation}}

\def\rhill{R_{\rm Hill}}

\def\mplanet{M_{\rm planet}}

\def\rhill{R_{\rm H}}

\def\simgr{\,\hbox{\hbox{$ > $}\kern -0.8em \lower 1.0ex\hbox{$\sim$}}\,}
\def\simle{\,\hbox{\hbox{$ < $}\kern -0.8em \lower 1.0ex\hbox{$\sim$}}\,}
\def\beq{\begin{equation}}
\def\eeq{\end{equation}}

\def\simgr{\,\hbox{\hbox{$ > $}\kern -0.8em \lower 1.0ex\hbox{$\sim$}}\,}
\def\simle{\,\hbox{\hbox{$ < $}\kern -0.8em \lower 1.0ex\hbox{$\sim$}}\,}
\def\beq{\begin{equation}}
\def\eeq{\end{equation}}

\def\({\left(}
\def\){\right)}
\def\<{\left<}
\def\>{\right>}

\def\({\left(} 
\def\){\right)} 
\def\<{\left<} 
\def\>{\right>} 

\def\bc{\begin{changebar}}
\def\bce{\begin{center}}
\def\beq{\begin{equation}} 

\def\bi{\begin{itemize}}

\def\btab{\begin{tabular}{p{1.7cm}p{12cm}p{1.5cm}}}
\def\bt2{\begin{tabular}{p{1 cm}p{4.5cm}p{10cm}}}

\def\ec{\end{changebar}}
\def\ece{\end{center}}
\def\eeq{\end{equation}} 
\def\ei{\end{itemize}}

\def\etab{\end{tabular}\\}

\def\mH2{m_\mathrm{H_2}}

\def\dmax0{\rho_\mathrm{max}}
\def\dmaxS0{\Sigma_\mathrm{max}}

\def\rH2{r_\mathrm{H_2}}

\def\r0max{r_\mathrm{0max}}

\def\s0{\sigma_\mathrm{0}}

\def\xp0{x_{\rm{M0}}}

\def\z0max{z_\mathrm{0max}}